\documentclass[twocolumn,showpacs,preprintnumbers,amsmath,amssymb]{revtex4}
\usepackage{graphicx}   % Include figure files
\usepackage{dcolumn}    % Align table columns on decimal point
\usepackage{bm}         % Bold math

\begin{document}
\title{Modularity and Extreme Edges of the Internet}

\author{Kasper Astrup Eriksen}
\email{kasper@nordita.dk}
\author{Ingve Simonsen}
\email{ingves@nordita.dk}
\affiliation{
The Nordic Institute for Theoretical Physics ---  NORDITA\\
Blegdamsvej 17, DK-2100 Copenhagen {\O}, Denmark}

\author{Sergei Maslov}
\email{maslov@bnl.gov}
\affiliation{Department of Physics, Brookhaven National Laboratory,
         Upton, New York 11973,  USA}

\author{Kim Sneppen}
\email{sneppen@nbi.dk}
\affiliation{The Niels Bohr Institute,
Blegdamsvej 17, DK-2100 Copenhagen {\O}, Denmark}
%\email{sneppen@phys.ntnu.no}
\affiliation{ Department of Physics, The Norwegian University of
Science and Technology, N-7491, Trondheim, Norway}

\date{\today}

%----------------------------------------------------------------------
%  ABSTRACT
%----------------------------------------------------------------------
\begin{abstract}
We study the spectral properties of a
diffusion process taking place on the Internet network
focusing on the slowest decaying modes. These modes
allow us to identify an underlying modular structure of the Internet
roughly corresponding to individual countries.
For instance in the slowest decaying mode the diffusion current
flows from Russia towards US military sites. These two regions
thus constitute the extreme edges of the Internet.
Quantitatively the modular structure of the Internet manifests
itself in approximately 10 times larger participation ratio of its
slow decaying modes compared to the null model -- a
random scale-free network.
We propose to use the fraction of nodes participating in
slow decaying modes as a general
measure of the modularity of a network. For the 100 slowest
decaying modes of the Internet we measured
this fraction to be around 30\%.
Finally we suggest, that the degree of isolation of an
individual module can be
assessed by comparing its participation in
different diffusion modes. Using the proportionality
of response as a criterion we find that
the independent module approximation works well for the
Internet.

\end{abstract}

%\keywords{Suggested keywords} % Use showkeys class option if keyword
                               % display desired
\maketitle

Virtually any complex system has
an underlying network that defines the backbone
of interactions among its components. Examples of such networks
include the Internet and the World Wide Web, molecular networks
of living cells, food webs in ecosystems, etc.
An important question is whether nodes of
such a network can be divided into smaller sub-networks (modules),
which interact with each other relatively weakly \cite{Girvan2002}.
Estimating the strength of inter-modular interactions,
localizing crucial links connecting these modules to each other,
and finding pairs of modules which are the most distant from
each other is important for several reasons.
First of all, it serves as a test of stability of the system
with respect to breaking it up into truly isolated components.
Such a break-up would be undesirable in, for example, the Internet.
Creation of extra connections between the most distant modules
in the network and reinforcement of crucial links is an efficient way
to increase its stability. Secondly, by measuring the relative strength of inter- and
intra-modular connections one directly assesses the quality of
the independent module approximation, which may turn out to be
important in modeling the actual dynamics of a given complex system.

In this work we explore the modular structure present in the
physical layout of the Internet. To this end we study
an auxiliary diffusion process taking place on this network.
The slowest modes of diffusion, easily identifiable
from the spectrum of its transfer matrix,
allow us to detect the weakly interacting modules
of the Internet.
These modules turn out to
roughly correspond to individual countries or for large countries
to cultural or geographical regions within the country.
Of course, the diffusion process studied
in this work does not describe the real dynamics of
the information flow over the Internet.
However, the detected modular features
play an important role in any local dynamical process
taking place on this network
including the real Internet traffic.

Analysis of spectral properties of
a similar diffusion process lies at the heart of the popular
search engine www.google.com \cite{google}. Its variants
have also been applied to social networks
(the correspondence analysis)
\cite{wasserman_book}, random and small-world networks
(the Laplace equation analysis) \cite{monasson},
artificial scale-free networks, \cite{FarkasBarabasiSpectra,GohKahngKimSpectra}
and the community structure of the World Wide Web
\cite{Kleinberg}.

In this work we explore the physical layout of the Internet
on a coarse-grained level of the so-called Autonomous Systems
(AS), which are large groups of routers and servers belonging to one
organization such as a university or a business
enterprise (e.g.\ an Internet Service Provider).
To this end we use the January 3, 2000 dataset
when the Internet consisted of 6474
Autonomous Systems exchanging information via 12572 undirected
links \cite{AS}. As expected for the Internet, any pair of
Autonomous Systems is connected to each other by at least one path,
so that topologically the network consists of just
one large cluster. The diffusion process we analyze here
describes the dynamics of a large number of
random walkers moving on the network
at discrete time steps. Statistical properties
of returns to the origin of such random walks
have recently been used to measure the effective
dimensionality of several complex networks \cite{Bilke2002}.
At each time-step every walker moves from its current node
to one of the neighboring nodes along a randomly selected
link.  The average dynamics of this process is described by
\begin{equation}
\rho_i (t+1) \; = \;  \sum_j \; T_{ij} \; \rho_j (t) \qquad ,
\label{transfer}
\end{equation}
where $\rho_i (t)$ is the expectation value of the
number of random walkers at site $i$ and time $t$.
The elements $T_{ij}$ of
the transfer matrix are equal to
$1/K_j$ for neighboring nodes $i$ and $j$
and zero otherwise. Here
$K_j$ is the connectivity (the number of immediate
neighbors) of the node $j$ from which a walker steps
to the node $i$.
Note that
%%the column sums of the transfer matrix are equal to one:
$\sum_i T_{ij}=1$,
so that the total number of walkers is conserved at all times.
Eq.~(\ref{transfer}) can also be rewritten as a discrete time
diffusion equation $\rho_i (t+1)- \rho_i (t) \; = \;  \sum_j \;
(T_{ij}-\delta_{ij}) \; \rho_j (t)$. Hence the diffusion
matrix $\mathbf{D}$ is related to the transfer matrix $\mathbf{T}$
simply as
\begin{equation}
\mathbf{D} = \mathbf{T} - \mathbf{1} \qquad .
\end{equation}
As time advances the distribution of
random walkers approaches a steady state $\rho_i(\infty)$
in which the diffusion current flowing from a node $i$ to a node $j$
is exactly balanced by that flowing from $j$ to $i$.
This is satisfied when the average number
of walkers $\rho_i(\infty)$ on every node $i$ is proportional
to its connectivity $K_i$.

The relaxation of any initial distribution of random walkers
among nodes, $\rho_i(0)$, towards the steady state configuration
 $\rho_i(\infty)$ is determined by the spectral properties of the matrix $\mathbf{T}$
\cite{CommentSymT} (or
alternatively $\mathbf{D}$). For instance, the steady state configuration
$\rho_i (\infty)$
itself is proportional to the principal eigenvector
$\rho_i^{(1)}$ of $\mathbf{T}$ corresponding
to its largest eigenvalue $\lambda^{(1)}=1$, which
is unique for single component networks such as the Internet.
The remaining eigenvectors $\rho^{(\alpha)}$
describe the decay of the initial configuration towards the steady state
with a characteristic decay time $\tau^{(\alpha)}$ related to the
corresponding eigenvalue $\lambda^{(\alpha)}$ through
$\exp (-1/\tau^{(\alpha)})=|\lambda^{(\alpha)}|$.
Note that in general there exist both non-oscillatory
($\lambda^{(\alpha)} \simeq 1$) and oscillatory
($\lambda^{(\alpha)} \simeq -1$) slowly decaying modes.

The modularity of a given
complex network reflects itself in statistical properties
of its diffusion eigenvectors $\rho_i^{(\alpha)}$.
One such property is the Participation Ratio (PR),
which quantifies the
effective number of nodes participating in a given eigenvector
with a significant weight.
In the Internet the components of the principal eigenvector
$\rho_i^{(1)} \propto \rho_i (\infty) \propto K_i$ as well as those of other
slow decaying modes are broadly distributed (scale-free)
\cite{Faloustos} and as such tend to be localized on just a few
highly connected nodes. In this case participation
ratios are best calculated using the normalized eigenvector
\begin{equation}
c_i^{(\alpha)} \; = \; \rho_i^{(\alpha)}/K_i
\label{currentdef}
\end{equation}
of outgoing currents flowing from node $i$ along each of its links.
More formally $c_i^{(\alpha)}$ is also the
eigenvector of the transposed transfer matrix
$\mathbf{T}^{\dagger}$ with the same eigenvalue
$\lambda^{(\alpha)}$. For such a vector normalized by
$\sum c_i^2=1$ the participation
ratio is defined as:
$PR \; = \; \left(\sum_{i=1}^N \; c_i^4\right)^{-1}$.

\begin{figure}[htbp]
  \centering
  \includegraphics*[width=7cm]{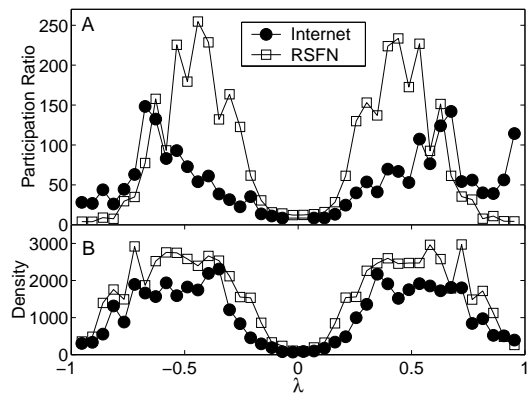}
  \caption{The participation ratio $\text{PR}^{(\alpha)}$ (top, A) and
the eigenvalue density
(bottom, B) as a function of the eigenvalue $-1<\lambda^{(\alpha)}<1$
measured in the Internet (filled circles) and in its randomized counterpart
(open squares) -- a Random
Scale-Free Network (RSFN). The participation ratio
was averaged over $\lambda$-bins of size 0.05 excluding
eigenmodes $\lambda^{(\alpha)}=0$ \cite{CommentDip1},
and $\lambda^{(1)}=1$.
Notice that for $|\lambda| \simeq 1$
participation ratios in the Internet
significantly exceed those in an RSFN indicating
the modular character of the former network.
}
\label{fig:ev}
\end{figure}

In Fig.~\ref{fig:ev}
the participation ratio of eigenvectors  $c_i^{(\alpha)}$
(top) and the eigenvalue density (i.e.\ the spectrum of the matrix)
(bottom) is plotted as a function of the corresponding
eigenvalue $-1<\lambda_i^{(\alpha)}<1$.
The data for
the Internet (filled circles) is displayed together
with the data for its randomized counterpart (open squares).
The randomization of the Internet was performed in such a way that the
connectivity of every node is strictly preserved \cite{maslov_sneppen_internet}.
It was argued \cite{maslov_sneppen_science,maslov_sneppen_internet} that such
a network constitutes a proper null model of a given complex network.
Since this random network has the same scale-free distribution of
connectivities as the Internet \cite{Faloustos} it will be referred to as
a Random Scale-Free Network (RSFN).

Comparing the data for the Internet and an RSFN we note that while the density
of states is rather similar in these two networks (Fig.~\ref{fig:ev}~B ),
the participation ratio of the slowly decaying modes (especially for the
non-oscillatory ones with $\lambda$ close to 1) is markedly higher in the Internet
than in an RSFN (Fig.~\ref{fig:ev}~A).
In these non-oscillatory modes the diffusion current flows from relatively
isolated regions (modules) along the few links connecting
them to the rest of the network.
If for such a module these links would be hypothetically
cut one by one, the corresponding eigenvalue
would gradually increase towards unity,
$\lambda^{(\alpha)} \rightarrow 1$,  while the
eigenvector would become more and more localized on the module.
When finally the module is completely disconnected
from the network the eigenvector has evolved to
the steady state solution on the module,
which has the participation ratio
equal to its size.
Thus the PR of slowly decaying
eigenmodes serves as a good quantitative estimate of the
size of modules in the network.
In an RSFN these modules are small
consisting of just a handful of nodes that
accidentally happen to be loosely connected to the rest of the network.
The fact that the participation ratios of slow decaying modes on the Internet
significantly exceed those in an RSFN indicates that the
corresponding modules are real and not accidental.
The average participation ratio of
slowly decaying modes can be quantified by
$\sum_{\alpha} \text{PR}^{({\alpha})} |\lambda^{({\alpha})}|^k/
\sum_{\alpha} |\lambda^{({\alpha})}|^k$.
For $5\le k\le 10$ this average changes only slowly
in both the Internet and an RSFN and equals approximately
$60$ and $5$, respectively.
A rough estimate for the number of different modules is given by the number of slowly decaying
non-oscillatory states
in Fig.~\ref{fig:ev}~A that have a participation ratio significantly exceeding
that of an RSFN. For the Internet the number is around $100$.
The sum of the participation ratios
for these first 100 modes,  $\sim 5400$ , is a rough estimate of the total
number of nodes in the modular part of the network.
This should be compared to the same sum being approximately equal to $520$
in an RSFN.
If one takes special care to avoid double counting nodes that appear more
than once among the set of $PR^{(\alpha)}$ nodes with the
largest $|c^{(\alpha)}_i|$ taken for each eigenmode $1 < \alpha \le 100$,
this number gets reduced to $\sim 1800$. Thus the overall modularity of the
Internet network is at least $1800/6500 \simeq 30\%$.

\begin{figure}[htbp]
  \centering
  \includegraphics*[width=7cm]{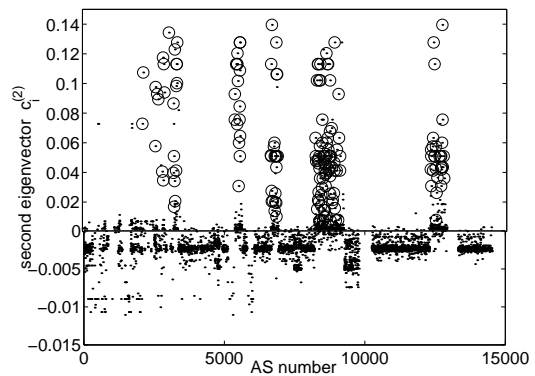}
  \caption{Components $c_i^{(2)}$ of the slowest decaying diffusion mode in the
  Internet (eigenvalue $\lambda^{(2)}= 0.9626$) as a function of the AS
  number. The AS known to be geographically located in Russia are
  marked with circles. The scale of the negative part of the
  $y$-axis is increased for clarity. Out of 100 Autonomous Systems with the
  most negative components $c_i^{(2)}$, those 23 for which we were
  able to find  the description are associated with the US Military.\
  }
  \label{fig:ASCur}
\end{figure}

To determine the organizing principle behind these Internet modules
in Fig.~\ref{fig:ASCur} we plot the outgoing current $c_i^{(2)}$
in the slowest decaying diffusion mode ($\lambda^{(2)}=0.9626$)
as a function of the AS number (note that some AS numbers are
not yet in use).
Autonomous Systems known to be located in Russia are marked with a circle.
The PR for this eigenmode is $~107$, while the total number
of Russian AS in our dataset is 174.
In Fig.~\ref{fig:ASCur} one can see
that almost all the Autonomous Systems
that significantly participate in this mode (large positive $c_i^{(2)}$) are Russian.
We have checked that the few exceptions to this rule are in fact Autonomous Systems
closely related to Russia.
Thus in the slowest decaying mode the diffusion current flows from a module that may be identified
with Russia towards the rest of the Internet. Curiously enough the set of Autonomous Systems
furthest away from Russia (the most negative $c_i^{(2)}$) are
located in the US and belong to the US Military.
This possible legacy of the cold war makes Russia and the US Military
the extreme edges of the Internet.
Performing a similar analysis for other slowly decaying modes we get a similar picture,
just with other pairs of countries being pulled out. For the Internet the
modules thus correspond to individual countries, or for large
countries -- to organizational or geographical features within the country.

It is interesting to note that these country-modules cannot be
detected using the spectral analysis of the adjacency matrix of
the network
\cite{SpectraOfGraphs,FarkasBarabasiSpectra,GohKahngKimSpectra}.
The elements of this matrix, closely related to $\mathbf{T}$, are
equal to 1 for a pair of neighboring nodes and 0 otherwise. The
largest eigenvectors of the adjacency matrix are known to be
localized primarily on the highest connected hubs and their
neighbors \cite{FarkasBarabasiSpectra,GohKahngKimSpectra}.
However, unlike in the case of $\mathbf{T}$, this undesirable localization
cannot be properly eliminated simply by dividing the components of
the eigenvectors by the connectivity $K_i$.
Hence eigenvectors of the adjacency matrix do not properly reflect the
country-based modular structure uncovered in this work.

\begin{figure}[htbp]
  \centering
  \includegraphics*[width=7cm]{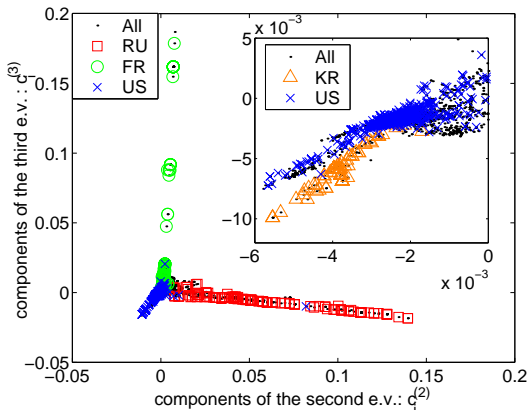}
  \caption{The Internet clustering:
Coordinate of the $i$-th AS in this plot are its components
$(c_i^{(2)},c_i^{(3)})$ in the two slowest decaying non-oscillatory
diffusion modes. The color code reveals the geographical location of the
AS: Russia - red squares, France - green circles, USA - blue crosses,
Korea - orange triangles). Note the straight lines corresponding
to good country-modules.
}
  \label{fig:internet}
\end{figure}

Having established that the Internet is indeed modular we now
address the question of how good these individual modules are.
To this end we compare different eigenmodes
$c_i^{(\alpha)}$ to each other.
Although the primary feature in a slowly decaying eigenmode
$c_i^{(\alpha)}$ is the flow between a dominant pair of
country-modules (such as between Russia and the US Military
in $c_i^{(2)}$), other modules may also
participate in it but to a smaller extent. This
gives rise to a fine structure within slow decaying modes
that is not captured by the participation ratio.
The hallmark of a good module
is that even though it participates in
different eigenmodes $\alpha$ to a different extent, the
relative distribution of $\alpha$-currents
within the module stays approximately the same.
In other words it enters different eigenmodes as just
one degree of freedom. In this case the ratio $c_i^{(\alpha)}/c_j^{(\alpha)}$
is approximately independent of the eigenmode $\alpha$
for any pair of nodes $i$ and $j$ within the module.
This is equivalent to the condition that for any two different eigenmodes
$\alpha$ and $\beta$,
$c_i^{(\alpha)}/c_i^{(\beta)}=\text{const }$ for every node $i$ belonging to the
given module.
In Fig.~\ref{fig:internet} we plot the outgoing currents in the two slowest decaying non-oscillatory
eigenmodes -- $c_i^{(2)}$, and $c_i^{(3)}$ -- as a function of each other.
Similar plots can be made for other pairs of slow decaying eigenmodes.
The principal feature in this kind of plot is a star like shape,
where different rays of the star
correspond to individual country-modules.
This type of plot is more powerful in identifying
individual modules than the participation ratio alone.
Indeed, in Fig.~\ref{fig:internet} one can easily detect
not only the most excited modules like
Russia, France, and US (red squares, green circles, and blue crosses),
but also less excited ones like Korea (orange triangles).
We believe that the idea of measuring the quality of
individual modules by how proportionally their nodes
participate in different slowly
decaying modes, can be easily generalized to other
dynamical processes taking place on the network
such as e.g. spin dynamics, vibrational modes, etc.

Finally, we would like to point out another interesting
feature of Fig.~\ref{fig:ev}. Both the density of states and the participation ratio
are nearly symmetric around $\lambda=0$ for both the Internet and an RSFN.
This near symmetry indicates
that both these networks are almost bipartite \cite{Comment1},
a feature also observed in citation networks, but not in
metabolic networks \cite{Bilke2002}.
In fact, a more detailed analysis shows that while almost every
slow oscillatory mode ($\lambda^{(\mathrm{osc})} \simeq -1$)
is related to the corresponding non-oscillatory mode with
$\lambda \simeq |\lambda^{(\mathrm{osc})}|$, the reverse is not
true as there are roughly 30\% more modes near $\lambda=1$ than near
$\lambda=-1$.
Those country-modules that are present in both the oscillatory and non-oscillatory
parts of the spectrum are internally almost bipartite. The simplest
bipartite graph is a tree and this seems to be the dominant structure
within the Internet modules.

In summa, we have demonstrated how a diffusion process taking place on
the Internet network allows one to extract information about its modules
and extreme edges. For many ``real-world'' complex networks the local context
of a node (in terms of the linkage pattern) reflects, or, perhaps, even
determines the importance and function of the given node.
For instance,
in biology one can successfully assign putative functions to
unclassified proteins based on the function of their interaction partners
\cite{SchwikowskiUetz}.
In general the diffusion process introduced in this work
can be seen as a systematic way to explore the local linkage structure
of a network beyond just the nearest neighbors. The detection of the
modular structure of a network is just one possible application
of such a process \cite{google,Bilke2002}.

Work at Brookhaven National Laboratory was carried out under Contract
No. DE-AC02-98CH10886, Division of Material Science, U.S.\ Department of
Energy. Two of us (K.E and K.S.) thank the Institute for Strongly
Correlated and Complex Systems at Brookhaven National Laboratory
for financial support during visits when part of this work was
completed.

% --------------------------------------------------------------------
% BIBLIOGRAPHY
% --------------------------------------------------------------------


\begin{references}
\bibitem{CommentSymT}
Even though $T_{ij}$ is not symmetric, all
of its eigenvalues and eigenvectors are real, since it
is related by a similarity transformation
to the symmetric matrix $\tilde{T}_{ij}=
T_{ij}\sqrt{K_j/K_i}$.
$\tilde{T}_{ij}$ has exactly the same
set of eigenvalues $\lambda^{(\alpha)}$ as $T_{ij}$,
and the corresponding eigenvectors $\tilde{\rho}_i^{(\alpha)}$
are related to the eigenvectors $\rho^{(\alpha)}_i$ of $T_{ij}$ via
$\tilde{\rho}^{(\alpha)}_i=\rho^{(\alpha)}_i/\sqrt{K_i}$.
\bibitem{CommentDip1}
Highly degenerate zero-modes ($\lambda^{(\alpha)}=0$) of a matrix
do not have a well defined PR. These modes are in fact rather
abundant and constitute $\approx 66\%$ of all the eigenvalues in
the Internet and $\approx 50\%$ in an RSFN. Zero-modes come
in part from pairs of nodes of connectivity 1 interacting with the
same highly connected partner. This structure is known to be
overrepresented in the
real Internet compared to an RSFN \cite{maslov_sneppen_internet}.
More generally zero-modes are localized on small substructures
in the network where every second node is of low connectivity.
\bibitem{Comment1}
In fact for a bipartite network
the spectrum is exactly symmetric and eigenvectors with eigenvalues
$\pm \lambda$ are related to each other by changing sign on all
nodes of one of the bipartite components.
\end{references}
\end{document}